\documentclass{svjour3}

\smartqed

\usepackage{graphicx}
\usepackage{amsfonts}
\usepackage{mathrsfs}
\usepackage{amsmath}
\usepackage{epstopdf}
\usepackage{hyperref}
\begin{document}

\begin{center}
{\large {\bf {Covariant formalism for the Berry connection due to gravity}}}
\end{center}

Achal Kumar\footnote{E-mail: achalkumar@iisc.ac.in} 
Banibrata Mukhopadhyay\footnote{E-mail: bm@iisc.ac.in}

{\it Department of Physics, Indian Institute of Science, Bangalore 560012, India}

\vspace{1.0cm}

\noindent
{\it Abstract:} 
It is well-known that Dirac particles gain geometric phase, namely Berry phase, while
moving in an electromagnetic field. Researchers have already shown covariant
formalism for the Berry connection due to an electromagnetic field. A similar effect is expected to happen due to the presence of Gravity. We use WKB
approximation to develop a covariant formalism of Berry-like connection in
the presence of Einstein gravity, which can be further used to describe the Berry-like phase or simply Berry phase. We also extend this formalism for massless Dirac particles (Weyl particles).Then we further show that this connection can be split
into two parts, one of which vanishes when the metric is spherically symmetric
and thus can be linked to the Aharonov-Bohm-like effect in the 3 + 1 formalism. At the same time, the other term can be related to the Pancharatnam-Berry like effect.  \\

\section*{1. Introduction}

Berry phase is a well-known geometric phase that has been adequately described 
in the evolution of spinors in the presence of an electromagnetic field. We expect that gravity should also cause a similar effect. Indeed we use WKB approximation for Dirac and Weyl particles to describe a covariant formalism for the Berry-like connection which can be further used to define Berry curvature, and to describe the evolution of the Berry-like phase of a particle as it moves in the presence of gravity. We perform the calculation of this covariant Berry connection for Dirac particles (massive spin-half) and Weyl particles (massless spin-half). 

The derivation we perform is based on a previous work [1], where Stone used WKB approximation in the Dirac equation to get the expression for Berry Connection in the presence of an electromagnetic field. We have a similar approach, but we start from the Dirac equation in curved spacetime, where the derivative is replaced by the spinorial covariant derivative. We see that this leads to a significant change in the equations that we get after using the WKB approximation. 

Once we have the expression for Berry connection due to gravitational effect, we would see that it can be split into two parts, one of which would vanish in the case of spherical symmetry. We will see that these terms can be defined as Pancharatnam-Berry-like (retained already) and Aharonov-Bohm-like (vanished in 
spherical symmetry), linking it to the 3 + 1 formalism of the geometric phase. Earlier it was found [2-3] that in 3 + 1 formalism, during the evolution of 
spinors in the presence of gravity, we obtain a Pancharatnam-Berry-like and Aharonov-Bohm-like term. Thus we expect the covariant formalism to show similar behavior, which we show through our calculations.
\section*{2. Dirac Equation in curved spacetime}
We start with the Dirac equation in curved spacetime. A Dirac field $\Psi(x)$ with a mass $m$ satisfies

\begin{equation}
	\big(i\hbar\gamma^\mu(\nabla_{\mu}) -m \big)\Psi(x) = 0,
\end{equation}
where $\gamma^{\mu}$-s are the spacetime gamma-matrices which are related to the flat space-time gamma-matrices by $ \gamma ^{\alpha} = e_{a}^{\alpha}\gamma^{a}$, $e_{a}^{\alpha}$ is the tetrad, the Greek and Latin indices imply 
repestively curved and flat coordinates, and $\nabla_{\mu}$ is the covariant derivative of spinors which is defined by a spin connection. The spinor derivative acts on the spinor field as 
\begin{equation}
	\nabla_{\mu}\Psi(x) = (\partial_{\mu} + \Omega_{\mu})\Psi(x),
\end{equation}
where $\Omega_{\mu}$ is defined as 
\begin{equation}
	\Omega_{\mu}= -\frac{i}{4}\omega_{ab\mu}(x)\sigma^{ab} = \frac{1}{8}\omega_{ab\mu}(x)[\gamma^{a},\gamma^{b}].
\end{equation}
Now we use the WKB ansatz 
\begin{equation}
	\Psi(x) = ae^{-\frac{i \varphi}{\hbar}},
\end{equation}
where $ a = a_{0} + \hbar a_{1} + \hbar^2 a_{2} + ....$, also $ \partial_{\mu} \varphi = p_{\mu}=(E,-\textbf{p})$. We consider the equations until the order of 
$h^1$ and use them to reach to our covaraint Berry connection. 

\section*{3. Dirac Particles}
This is the case where $m \neq 0$. At the order of $\hbar^{0}$ we obtain
\begin{equation*}
	(\gamma^{\mu}p_{\mu} - m)a_{0} = 0. 
\end{equation*}
Considering $ u_{\alpha}$ to be the complete set of eigenspinor solutions of the above  equation such that
\begin{equation}
	(\gamma^{\mu}p_{\mu} - m)u_{\alpha} = 0, 
\end{equation}
$a_{0}= u_{\alpha}(p)C(x)^{\alpha}$ where $C^{\alpha}$ is the complex number. 
\newline
At the order of $h^{1}$, we then obtain
\begin{equation}
	i(\gamma^{\mu} \nabla_{\mu} a_0 ) + (\gamma^{\mu}p_{\mu} - m)a_1 = 0.  
\end{equation}
We would also need the relation for the four-current of Dirac particle which is 
\begin{equation}
	\bar{u}_{\beta}\gamma^{\mu} u_{\alpha} =\delta_{\beta \alpha}\frac{p^{\mu}}{m} =   \delta_{\beta \alpha}V^{\mu}.    
\end{equation}
Using Eqs. (5)-(7) and the completeness and normalization of $u_{\alpha}$, it can be shown that
\begin{equation}
	(\delta_{\alpha \beta } V^{\mu} \nabla_{\mu} + \frac{1}{2} \delta_{\alpha \beta} \nabla_{\mu} V^{\mu}   +     \frac{1}{2}V^{\mu}(\bar{u}_{\alpha}\nabla_{\mu}u_{\beta } - \nabla_{\mu}\bar{u}_{\alpha} u_{\beta} ) )C^{\alpha} = 0.    
\end{equation}
Here when we apply $\nabla_{\mu}$ to $V^{\mu}$ we are considering the covariant derivative for tensors which comprises of the Christoffel connection. We are using the same symbol $\nabla_{\mu}$ for both spnior covariant derivative and covariant derivative for tensor fields. This should not cause any confusion, since the type of covariant derivative can be
determined by the object on which it acts on.

Now we can define our covariant Berry connection as

\begin{equation}
	B_{\mu \alpha \beta } =\frac{i}{2}(\bar{u}_{\alpha}\nabla_{\mu}u_{\beta } - \nabla_{\mu}\bar{u}_{\alpha} u_{\beta} ), 
\end{equation}
which can be used to define berry curvature and to describe the Berry phase that a Dirac particle gains due to gravity. The details of the above derivation can be found elsewhere.

\section*{4. Weyl Particles}

In this case we consider spin half particles with mass $m=0$, Thus the Dirac equation for massless particles is 

\begin{equation}
	\gamma^\mu \nabla_{\mu}  \Psi(x) = 0.
\end{equation}
We use here Weyl representation for the gamma-matrices to make our calculations easier. We see that this case is significantly different than the last case because, first of all, the Dirac equation does not have $\hbar$, which we used as a parameter in the last case. Hence, we define a small parameter $\epsilon $ and expand our solution in terms of that. We further seek WKB solution and
choose

\begin{equation}
	\Psi(x) = ae^{-\frac{i \varphi}{\epsilon}},
\end{equation}
where $ a = a_{0} + \epsilon a_{1} + \epsilon^2 a_{2} + ....$, also $ \partial_{\mu} \varphi = p_{\mu}$ and $\epsilon$ is the small parameter. Similar to the 
case of Dirac particles, we substitute Eq. (11) in Eq. (10) and expand in the orders of $\epsilon$. Considering terms upto the order $\epsilon^0$, we have

\begin{equation*}
	\gamma^{\mu}a_{0}\Bigg ( \frac{-i \partial_{\mu}\varphi}{\epsilon}\Bigg )e^{-i\varphi/\epsilon} + \gamma^{\mu}(\nabla_{\mu}a_{0})e^{-i\varphi/\epsilon} + \gamma_{\mu}a_1\epsilon\Bigg( \frac{-i \partial_{\mu}\varphi}{\epsilon} \Bigg)e^{-i\varphi/\epsilon} = 0, 
\end{equation*}

\begin{equation*}
	\gamma^{\mu}a_{0}\Bigg ( \frac{-i p_{\mu}}{\epsilon}\Bigg ) + \gamma^{\mu}(\nabla_{\mu}a_{0}) + \gamma^{\mu}a_1( -i p_{\mu} ) = 0. 
\end{equation*}

At the order of $\epsilon^{-1}$ we obtain

\begin{equation}
	\gamma^{\mu}p_{\mu}a_{0} = 0. 
\end{equation}

Let $U_{L}$ and $ U_{R}$ be respectively the left-handed and right-handed eigenspinor solutions of Eq. (12) with positive energy $(p_{0}>0)$. Similarly $V_{L}$ and $V_{R}$ be respectively the left- and right-handed anti-particle eignespinors $(p_{0}<0)$. Thus we can say that $a_{0}= U_{\alpha}(p)C(x)^{\alpha}$ where $C^{\alpha}$ is the complex number and $\alpha \in \{L,R\}$.

At the order $\epsilon^{0}$ 
\begin{equation}
	\gamma^{\mu} \nabla_{\mu} a_0 - i\gamma^{\mu}p_{\mu}a_1 = 0.  
\end{equation}
\newline

Pre-multiplying Eq. (13) with $\bar{U}_{\beta}$ where $\beta \in \{L,R \} $ and using $ \bar{U}_{\beta}\gamma^{\mu}p_{\mu} = 0 $, we obtain

\begin{equation}
	\bar{U}_{\beta}(\gamma^{\mu} \nabla_{\mu}a_0 ) = 0.
\end{equation}
Similarly, expanding $a_0 = U_{\alpha } C^{\alpha}$, we obtain

\begin{equation}
	(\bar{U}_{\beta}\gamma^{\mu}U_{\alpha})\nabla_{\mu}C^{\alpha} + (\bar{U}_{\beta}\gamma^{\mu}\nabla_{\mu}U_{\alpha})C^{\alpha} = 0.
\end{equation}
Now we need to use a relation for the four-current to isolate the Berry connection. However, clearly, simply putting $m=0$ in the relation that we used in the massive case described in Section 3 would not work because the RHS diverges in (7). It is possible to derive the relation for the four-current if we use the two-spinor representation for the Eigen spinors. We know that the four-spinors can be written in terms of the two-spinors as

\begin{equation}
	U_{L} =  \begin{pmatrix}
		u_{L}\\
		0
	\end{pmatrix} \text{and } U_{R} =  \begin{pmatrix}
		0\\
		u_{R}
	\end{pmatrix},
\end{equation}
\begin{equation}
	\bar{U}_{L} = (
	0 , u_{L}^{\dag}) \text{ and } \bar{U}_{R} = (
	u_{R}^{\dag} , 0 ),
\end{equation}

\begin{equation}
	\gamma^{\mu} = 
	\begin{bmatrix}
		0 & \sigma^{\mu} \\
		\bar{\sigma}^{\mu} & 0 
	\end{bmatrix},
\end{equation}
where $\bar{\sigma}^{\mu} = (1,-\Vec{\sigma})$ and $\sigma^\mu = (1, \Vec{\sigma})$ and $u_{\alpha}$-s are two-spinors. Note that in Eqs. (16) and (17), 0 represents $\begin{pmatrix}
	0\\
	0
\end{pmatrix}$ and similarly in Eq. (18) it represents $2\times2$ null matrix. Same relations are valid for antiparticle four-spinors $V_{\alpha}$ and their 
corresponding two-spinors $v_{\alpha}$. 

Substituting these relations is Eq. (10) we can derive the relation for the 
four-current in this case. The details of this calculations can be found elsewhere. We will finally obtain that 

\begin{equation}
	\bar{U}_{\alpha}\gamma^{\mu} U_{\beta} = -\delta_{\alpha \beta} \frac{p^{\mu}}{p_{0}} = -\delta_{\alpha \beta}\frac{p^{\mu}}{p_{\nu}e_{0}^{\nu}} = \delta_{\alpha\beta}H^{\mu},
\end{equation}
where $ \alpha,\beta \in \{L,R\}$. We see that the quantity $H^{\mu}$ is frame-dependent because of the fact it contains $e^{\nu}_{0}$, which is different from the massive case. This is because of the fact that for a massless particle, there is no rest frame. This is the reason that the expression that we obtain for the four-current is also observer-dependent. Using Eqs. (19) and (15), it can be shown that 

\begin{equation}
	(\delta_{\alpha \beta } H^{\mu} \nabla_{\mu} + \frac{1}{2} \delta_{\alpha \beta} \nabla_{\mu} H^{\mu}   +     \frac{1}{2}H^{\mu}(\bar{U}_{\alpha}\gamma^0\nabla_{\mu}U_{\beta } - \nabla_{\mu}\bar{U}_{\alpha} \gamma^0 U_{\beta} ) )C^{\beta} = 0.    
\end{equation}
The details of these calculations can be found elsewhere.
Now similar to the massive case, we can define the Berry connection as 

\begin{equation}
	B_{\mu \alpha \beta } =\frac{i}{2}(\bar{U}_{\alpha}\gamma^0\nabla_{\mu}U_{\beta } - \nabla_{\mu}\bar{U}_{\alpha} \gamma^0 U_{\beta} ). 
\end{equation} 

Here we see that we have extra $\gamma^0$ in the expression when compared to the massive case  which is due to the fact that we are using Weyl's representation of gamma matrices.

\section*{5. Pancharatnam-Berry-like and Aharonov-Bohm-like terms}

Now in the massive case we would use Eq. (2) in Eq. (9) to isolate the Pancharatnam-Berry-like and Aharonov-Bohm-like terms. Taking adjoint of Eq. (2) we have

\begin{equation}
	\nabla_{\mu}u_{\alpha}^{\dagger} = \partial_{\mu} u_{\alpha}^{\dagger} + u_{\alpha}^{\dagger}(\Omega_{\mu})^{\dagger}.
\end{equation}                 
                               
It can be shown using Eq. (3) that $(\Omega_{\mu})^{\dagger} = - \gamma_{0}\Omega_{\mu}\gamma_{0}$. Thus using this and multiplying $\gamma_{0}$ from the right in Eq. (22) we obtain

\begin{equation}
	\nabla_{\mu}\bar{u}_{\alpha} = \partial_{\mu} \bar{u}_{\alpha} - \bar{u}_{\alpha}\Omega_{\mu}.
\end{equation}                 

Substituting Eqs. (22) and (23) in Eq. (9), we obtain

\begin{equation}
		B_{\mu \alpha \beta } =\frac{i}{2}\Big[(\bar{u}_{\alpha}\partial_{\mu}u_{\beta } - \partial_{\mu}\bar{u}_{\alpha} u_{\beta} )  + 2(\bar{u}_{\alpha}\Omega_{\mu}u_{\beta})\Big].
\end{equation}

Only the term $2(\bar{u}_{\alpha}\Omega_{\mu}u_{\beta})$ is due to the spinor connection. The rest of the terms are also present in the case of flat spacetime and correspond to the Berry connection in flat spacetime [1]. Thus we can say 

\begin{equation}
		B_{\mu \alpha \beta } =B^{flat}_{\mu \alpha \beta }  + i(\bar{u}_{\alpha}\Omega_{\mu}u_{\beta}).
\end{equation}

Now we can split this extra term to seperate the Pancharatnam-Berry-like and Aharonov-Bohm-like terms. We can expand the term

\begin{equation}
	\bar{u}_{\alpha}\Omega_{\nu}u_{\beta} = \frac{1}{8}(\omega_{ab\nu})(\bar{u}_{\alpha}[\gamma^{a},\gamma^{b}]u_{\beta}),
\end{equation}
Hence,
\begin{equation}
	\bar{u}_{\alpha}\Omega_{\nu}u_{\beta} =
	\frac{1}{8}(\omega_{0b\nu}\bar{u}_{\alpha}[\gamma^{0},\gamma^{b}]u_{\beta} + \omega_{a0\nu}\bar{u}_{\alpha}[\gamma^{a},\gamma^{0}]u_{\beta} + \bar{u}_{\alpha}\omega_{ij\nu}[\gamma^{i},\gamma^{j}]u_{\beta}),
\end{equation}

where $i , j \in \{1,2,3\}$, which can be simplified as 

\begin{equation}
	\bar{u}_{\alpha}\Omega_{\nu}u_{\beta} =
	\frac{1}{8}((\omega_{0i\nu} - \omega_{i0\nu})\bar{u}_{\alpha}[\gamma^0,\gamma^i]u_{\beta}) +  \frac{1}{8}\omega_{ij\nu}\bar{u}_{\alpha}[\gamma^{i},\gamma^{j}]u_{\beta}.
\end{equation}

Now the second term of R.H.S. vanishes when the metric is spherically symmetric. If the metric is spherically symmetric then $\omega_{ij\nu}$ is non zero only when $i=j$ but then $[\gamma^{i},\gamma^j] = 0$ thus making the term $\frac{1}{8}\omega_{ij\nu}\bar{u}_{\alpha}[\gamma^{i},\gamma^{j}]u_{\beta} = 0$. Therefore, we can define the first and second terms as Pancharatnam-Berry-like and Aharonov-Bohm-like respectively. Thus we have 
\begin{equation}
	B_{\mu \alpha \beta } =B^{flat}_{\mu \alpha \beta }  + 	\frac{i}{8}((\omega_{0i\nu} - \omega_{i0\nu})\bar{u}_{\alpha}[\gamma^0,\gamma^i]u_{\beta}) +  \frac{i}{8}\omega_{ij\nu}\bar{u}_{\alpha}[\gamma^{i},\gamma^{j}]u_{\beta}.
\end{equation}


\section*{6. Conclusion}

We have found the covariant description of Berry connection for spin half particles. The connection that we have derived is in the absence of an electromagnetic field. The same calculations can be repeated considering the electromagnetic field as well, and we would get a similar result, but the final expression would contain additional terms with $F_{\mu\nu}$. These calculations can be found elsewhere. We have found a similar expression for the Berry connection in the massless case. However, the connection there is an observer-dependent quantity, unlike the massive case. This can be attributed to the fact that there is no rest frame for a massless particle. Finally, we have found that it is possible to isolate the  Pancharatnam-Berry-like and Aharonov-Bohm-like terms from the Berry connection. The last result was also observed in the 3 + 1 formalism of the geometric phase [2-4]. 
\bibliographystyle{unsrt}

\end{document}